\documentstyle[amssymb,12pt,thmsa,sw20lart]{article}

\input{tcilatex}
\begin{document}

\title{Four-level quantum teleportation, swapping and collective translations of
multipartite quantum entanglement}
\author{Zai-Zhe Zhong \\
Department of Physics, Liaoning Normal University, Dalian 116029, \\
Liaoning,China. E-mail: zhongzaizhecn@yahoo.com.cn}
\maketitle

\begin{abstract}
In this paper, an optimal scheme of four-level quantum teleportation and
swapping of quantum entanglement is given. We construct a complete
orthogonal basis of the bipartite ququadrit systems. Using this basis, the
four-level quantum teleportation and swapping can be achieved according to
the standard steps. In addition, associate the above bases with the
unextendible product bases and the exact entanglement bases, we prove that
in the $2\times 2\times 2$ systems or $3\times 3$ systems the collective
translations of multipartite quantum entanglement can be realized.

PACC numbers: 03.67.Mn, 03.65.Ud, 03.67.Hk.

\end{abstract}

Quantum teleportation and swapping are quite interesting and important
topics in modern quantum mechanics and quantum information. Following the
BBCJPW scheme[1], there have been very many related works (e.g. see the
references in [2]). About the swapping of quantum entanglement, the original
work was by ZZHE[3$].$ About the problems how to extend the schemes of
quantum teleportation and swapping to the multipartite d-level ($d\geqslant
3)$ cases, there are many works (e.g. see [4-9], and some recent works
related to teleporting states of dimension higher than two, see [10,11]). In
this paper, we point out that in the case of ququadrit systems we can find a
simple way to achieve quantum teleportation and swapping. We give a new
complete orthogonal basis. Using this basis and according to the standard
steps, the four-level quantum teleportation and swapping can be realized.
The mathematical form of the schemes is simple.

In addition, there is yet another interesting result that we can achieve
collective quantum teleportation and collective swapping of quantum
entanglement of two or three particles. In this paper, as an application of
the above schemes, we prove that this is possible, i.e. we associate the
above basis with the unextendible product bases (UPB)[12-16] and the exact
entanglement bases (EEB)[17], so that in the $2\times 2\times 2$ systems or $%
3\times 3$ systems collective translations of multipartite quantum
entanglement can be realized.

If $H^{\left( 1\right) }$ and $H^{\left( 2\right) }$ are the Hilbert spaces
of ququadrit states with the natural bases $\left\{ \mid i\rangle \right\} $
and $\left\{ \mid j\rangle \right\} \left( i,j=0,1,2,3\right) ,$
respectively, then the natural basis of $H=H^{\left( 1\right) }\otimes
H^{\left( 2\right) }$ is $\left\{ \mid i\rangle \mid j\rangle \right\} .$ In
the first place, we define a new complete orthogonal basis $\left\{ \mid
W_i\rangle ,\mid X_i\rangle ,\mid Y_i\rangle ,\mid Z_i\rangle \right\}
\left( i=0,1,2,3\right) $ of $H$ by 
\begin{eqnarray}
&\mid &W_i\rangle =\frac 12\left( \mid i\rangle \mid 0\rangle +\mid i+1\;%
\func{mod}4\rangle \mid 1\rangle +\mid i+2\;\func{mod}4\rangle \mid 2\rangle
+\mid i+3\;\func{mod}4\rangle \mid 3\rangle \right)  \nonumber \\
&\mid &X_i\rangle =\frac 12\left( \mid i\rangle \mid 0\rangle +\mid i+1\;%
\func{mod}4\rangle \mid 1\rangle -\mid i+2\;\func{mod}4\rangle \mid 2\rangle
-\mid i+3\;\func{mod}4\rangle \mid 3\rangle \right) \\
&\mid &Y_i\rangle =\frac 12\left( \mid i\rangle \mid 0\rangle -\mid i+1\;%
\func{mod}4\rangle \mid 1\rangle +\mid i+2\;\func{mod}4\rangle \mid 2\rangle
-\mid i+3\;\func{mod}4\mid 3\rangle \right)  \nonumber \\
&\mid &Z_i\rangle =\frac 12\left( \mid i\rangle \mid 0\rangle -\mid i+1\;%
\func{mod}4\rangle \mid 1\rangle -\mid i+2\;\func{mod}4\rangle \mid 2\rangle
+\mid i+3\;\func{mod}4\rangle \mid 3\rangle \right)  \nonumber
\end{eqnarray}
Obviously, $\left\{ \mid W_i\rangle ,\mid X_i\rangle ,\mid Y_i\rangle ,\mid
Z_i\rangle \right\} \left( i=0,1,2,3\right) $ are entangled states, they
form a complete orthogonal basis of $H$. The transformation relation between
the above basis and the natural basis is 
\begin{eqnarray}
&\mid &i\rangle \mid 0\rangle =\frac 12\left( \mid W_i\rangle +\mid
X_i\rangle +\mid Y_i\rangle +\mid Z_i\rangle \right)  \nonumber \\
&\mid &i+1\;\func{mod}4\rangle \mid 1\rangle =\frac 12\left( \mid W_i\rangle
+\mid X_i\rangle -\mid Y_i\rangle -\mid Z_i\rangle \right)  \nonumber \\
&\mid &i+2\;\func{mod}4\rangle \mid 2\rangle =\frac 12\left( \mid W_i\rangle
-\mid X_i\rangle +\mid Y_i\rangle -\mid Z_i\rangle \right) \\
&\mid &i+3\;\func{mod}4\rangle \mid 3\rangle =\frac 12\left( \mid W_i\rangle
-\mid X_i\rangle -\mid Y_i\rangle +\mid Z_i\rangle \right) \   \nonumber
\end{eqnarray}

Following the standard steps of quantum teleportation, we suppose that Alice
holds particle 1 which is in an unknown pure-state $\mid \phi ^{\left(
1\right) }\rangle =\alpha \mid 0_1\rangle +\beta \mid 1_1\rangle +\gamma
\mid 2_1\rangle +\delta \mid 3_1\rangle ,$ Clara is in a remote place from
Alice. Bob holds two particles 2 and 3 which are in some basic state, for
instance, in $\mid X_1^{\left( 2,3\right) }\rangle =\frac 12\left( \mid
1_2\rangle \mid 0_3\rangle +\mid 2_2\rangle \mid 1_3\rangle -\mid 3_2\rangle
\mid 2_3\rangle -\mid 0_2\rangle \mid 3_3\rangle \right) $. The total state
is 
\begin{equation}
\begin{array}{c}
\mid \Psi _{total}\rangle =\mid \phi ^{\left( 1\right) }\rangle \mid
X_1^{\left( 2,3\right) }\rangle =\frac 12(\alpha \mid 0_1\rangle \mid
1_2\rangle \mid 0_3\rangle +\alpha \mid 0_1\rangle \mid 2_2\rangle \mid
1_3\rangle -\alpha \mid 0_1\rangle \mid 3_2\rangle \mid 2_3\rangle \\ 
-\alpha \mid 0_1\rangle \mid 0_2\rangle \mid 3_3\rangle +\beta \mid
1_1\rangle \mid 1_2\rangle \mid 0_3\rangle +\cdots -\delta \mid 3_1\rangle
\mid 0_2\rangle \mid 3_3\rangle )
\end{array}
\end{equation}
According to Eq.(2), every $\mid i_1\rangle \mid j_2\rangle $ can always be
expressed by $\mid W_k^{\left( 1,2\right) }\rangle ,\;\mid X_k^{\left(
1,2\right) }\rangle ,\;$ $\mid Y_k^{\left( 1,2\right) }\rangle $ and $\mid
Z_k^{\left( 1,2\right) }\rangle .$ Substitute them into Eq.(3) and
reorganize, the final result is 
\begin{eqnarray}
&\mid &\Psi _{total}\rangle =\sum_{i=0}^3\left( 
\begin{array}{c}
\mid W_i^{\left( 1,2\right) }\rangle \mid \phi _{W_i}^{\left( 3\right)
}\rangle +\mid X_i^{\left( 1,2\right) }\rangle \mid \phi _{X_i}^{\left(
3\right) }\rangle \\ 
+\mid Y_i^{\left( 1,2\right) }\rangle \mid \phi _{Y_i}^{\left( 3\right)
}\rangle +\mid Z_i^{\left( 1,2\right) }\rangle \mid \phi _{Z_i}^{\left(
3\right) }\rangle
\end{array}
\right)  \nonumber \\
&\equiv &\frac 14\left\{ 
\begin{array}{c}
\mid W_0^{\left( 1,2\right) }\rangle \left( \beta \mid 0_3\rangle +\gamma
\mid 1_3\rangle -\delta \mid 2_3\rangle -\alpha \mid 3_3\rangle \right) \\ 
+\mid W_1^{\left( 1,2\right) }\rangle \left( \gamma \mid 0_3\rangle +\delta
\mid 1_3\rangle -\alpha \mid 2_3\rangle -\beta \mid 3_3\rangle \right) \\ 
+\mid W_2^{\left( 1,2\right) }\rangle \left( \delta \mid 0_3\rangle +\alpha
\mid 1_3\rangle -\beta \mid 2_3\rangle -\gamma \mid 3_3\rangle \right) \\ 
+\mid W_3^{\left( 1,2\right) }\rangle \left( \alpha \mid 0_3\rangle +\beta
\mid 1_3\rangle -\gamma \mid 2_3\rangle -\delta \mid 3_3\rangle \right) \ 
\\ 
\, \\ 
+\mid X_0^{\left( 1,2\right) }\rangle \left( \beta \mid 0_3\rangle -\gamma
\mid 1_3\rangle -\delta \mid 2_3\rangle -\alpha \mid 3_3\rangle \right) \\ 
+\mid X_1^{\left( 1,2\right) }\rangle \left( \gamma \mid 0_3\rangle -\delta
\mid 1_3\rangle +\alpha \mid 2_3\rangle -\beta \mid 3_3\rangle \right) \\ 
+\mid X_2^{\left( 1,2\right) }\rangle \left( \delta \mid 0_3\rangle -\alpha
\mid 1_3\rangle +\beta \mid 2_3\rangle -\gamma \mid 3_3\rangle \right) \\ 
+\mid X_3^{\left( 1,2\right) }\rangle \left( -\alpha \mid 0_3\rangle -\beta
\mid 1_3\rangle -\gamma \mid 2_3\rangle -\delta \mid 3_3\rangle \right) \ 
\\ 
\, \\ 
+\mid Y_0^{\left( 1,2\right) }\rangle \left( -\beta \mid 0_3\rangle +\gamma
\mid 1_3\rangle +\delta \mid 2_3\rangle -\alpha \mid 3_3\rangle \right) \\ 
+\mid Y_1^{\left( 1,2\right) }\rangle \left( -\gamma \mid 0_3\rangle +\delta
\mid 1_3\rangle +\alpha \mid 2_3\rangle -\beta \mid 3_3\rangle \right) \\ 
+\mid Y_2^{\left( 1,2\right) }\rangle \left( -\delta \mid 0_3\rangle -\alpha
\mid 1_3\rangle -\beta \mid 2_3\rangle -\gamma \mid 3_3\rangle \right) \\ 
+\mid Y_3^{\left( 1,2\right) }\rangle \left( -\alpha \mid 0_3\rangle +\beta
\mid 1_3\rangle +\gamma \mid 2_3\rangle -\delta \mid 3_3\rangle \right) \ 
\\ 
\, \\ 
+\mid Z_0^{\left( 1,2\right) }\rangle \left( -\beta \mid 0_3\rangle -\gamma
\mid 1_3\rangle -\delta \mid 2_3\rangle -\alpha \mid 3_3\rangle \right) \\ 
+\mid Z_1^{\left( 1,2\right) }\rangle \left( -\gamma \mid 0_3\rangle -\delta
\mid 1_3\rangle -\alpha \mid 2_3\rangle -\beta \mid 3_3\rangle \right) \\ 
+\mid Z_2^{\left( 1,2\right) }\rangle \left( -\delta \mid 0_3\rangle +\alpha
\mid 1_3\rangle +\beta \mid 2_3\rangle -\gamma \mid 3_3\rangle \right) \\ 
+\mid Z_3^{\left( 1,2\right) }\rangle \left( \alpha \mid 0_3\rangle -\beta
\mid 1_3\rangle -\gamma \mid 2_3\rangle -\delta \mid 3_3\rangle \right)
\end{array}
\right\}
\end{eqnarray}
Now, Bob sends particles 2 and 3 to Alice and Clara respectively, and Alice
makes a joint measurement of particles 1, 2. She will obtain of the 16 basic
states $\left\{ \mid W_i^{\left( 1,2\right) }\rangle ,\mid X_i^{\left(
1,2\right) }\rangle ,\mid Y_i^{\left( 1,2\right) }\rangle ,\mid Z_i^{\left(
1,2\right) }\rangle \right\} \left( i=0,1,2,3\right) $ with probability $%
\frac 1{16}$ (here we assume that there is such instrument which can
distinguish these bases$)$. Simultaneously Clara must obtain a corresponding
one state of $\left\{ \mid \phi _{W_i}^{\left( 3\right) }\rangle ,\mid \phi
_{X_i}^{\left( 3\right) }\rangle ,\mid \phi _{Y_i}^{\left( 3\right) }\rangle
,\mid \phi _{Z_i}^{\left( 3\right) }\right\} .$ We define sixteen unitary
matrices as 
\begin{eqnarray}
U_{W_0} &=&\left[ 
\begin{array}{llll}
0 & 0 & 0 & -1 \\ 
1 & 0 & 0 & 0 \\ 
0 & 1 & 0 & 0 \\ 
0 & 0 & -1 & 0
\end{array}
\right] ,\;U_{W_1}=\left[ 
\begin{array}{llll}
0 & 0 & -1 & 0 \\ 
0 & 0 & 0 & -1 \\ 
1 & 0 & 0 & 0 \\ 
0 & 1 & 0 & 0
\end{array}
\right]  \nonumber \\
\;U_{W_2} &=&\left[ 
\begin{array}{llll}
0 & 1 & 0 & 0 \\ 
0 & 0 & -1 & 0 \\ 
0 & 0 & 0 & -1 \\ 
1 & 0 & 0 & 0
\end{array}
\right] ,\;U_{W_3}=\left[ 
\begin{array}{llll}
1 & 0 & 0 & 0 \\ 
0 & 1 & 0 & 0 \\ 
0 & 0 & -1 & 0 \\ 
0 & 0 & 0 & -1
\end{array}
\right]  \nonumber \\
&& \\
U_{X_0} &=&\left[ 
\begin{array}{llll}
0 & 0 & 0 & -1 \\ 
1 & 0 & 0 & 0 \\ 
0 & -1 & 0 & 0 \\ 
0 & 0 & -1 & 0
\end{array}
\right] ,\;U_{X_1}=\left[ 
\begin{array}{llll}
0 & 0 & 1 & 0 \\ 
0 & 0 & 0 & -1 \\ 
1 & 0 & 0 & 0 \\ 
0 & -1 & 0 & 0
\end{array}
\right]  \nonumber \\
\;U_{X_2} &=&\left[ 
\begin{array}{llll}
0 & -1 & 0 & 0 \\ 
0 & 0 & 1 & 0 \\ 
0 & 0 & 0 & -1 \\ 
1 & 0 & 0 & 0
\end{array}
\right] ,\;U_{X_3}=\left[ 
\begin{array}{llll}
-1 & 0 & 0 & 0 \\ 
0 & -1 & 0 & 0 \\ 
0 & 0 & -1 & 0 \\ 
0 & 0 & 0 & -1
\end{array}
\right]  \nonumber \\
&& \\
U_{Y_0} &=&\left[ 
\begin{array}{llll}
0 & 0 & 0 & -1 \\ 
-1 & 0 & 0 & 0 \\ 
0 & 1 & 0 & 0 \\ 
0 & 0 & 1 & 0
\end{array}
\right] ,\;U_{Y_1}=\left[ 
\begin{array}{llll}
0 & 0 & 1 & 0 \\ 
0 & 0 & 0 & -1 \\ 
-1 & 0 & 0 & 0 \\ 
0 & 1 & 0 & 0
\end{array}
\right] \\
\;U_{Y_2} &=&\left[ 
\begin{array}{llll}
0 & -1 & 0 & 0 \\ 
0 & 0 & -1 & 0 \\ 
0 & 0 & 0 & -1 \\ 
-1 & 0 & 0 & 0
\end{array}
\right] ,\;U_{Y_3}=\left[ 
\begin{array}{llll}
-1 & 0 & 0 & 0 \\ 
0 & 1 & 0 & 0 \\ 
0 & 0 & 1 & 0 \\ 
0 & 0 & 0 & -1
\end{array}
\right]  \nonumber \\
&& \\
U_{Z_0} &=&\left[ 
\begin{array}{llll}
0 & 0 & 0 & -1 \\ 
-1 & 0 & 0 & 0 \\ 
0 & -1 & 0 & 0 \\ 
0 & 0 & -1 & 0
\end{array}
\right] ,\;U_{Z_1}=\left[ 
\begin{array}{llll}
0 & 0 & -1 & 0 \\ 
0 & 0 & 0 & -1 \\ 
-1 & 0 & 0 & 0 \\ 
0 & -1 & 0 & 0
\end{array}
\right]  \nonumber \\
\;U_{Z_2} &=&\left[ 
\begin{array}{llll}
0 & 1 & 0 & 0 \\ 
0 & 0 & 1 & 0 \\ 
0 & 0 & 0 & -1 \\ 
-1 & 0 & 0 & 0
\end{array}
\right] ,\;U_{Z_3}=\left[ 
\begin{array}{llll}
1 & 0 & 0 & 0 \\ 
0 & -1 & 0 & 0 \\ 
0 & 0 & -1 & 0 \\ 
0 & 0 & 0 & -1
\end{array}
\right]  \nonumber
\end{eqnarray}
When Alice informs Clara of her measurement result $\mu _j$, by the
classical communications, then Clara at once knows the correct result should
be $\mid \phi ^3\rangle =U_{\mu _j}\mid \phi _{\mu _j}^3\rangle $. Thus we
achieve a four-level quantum teleportation.

By using of the above basis, we can also carry out the four-level swapping.
For instance, we suppose that Alice holds particle 1, Bob holds particles 2,
3 , Clara holds particle 4, particles 1 and 2 are in the entangled state $%
\mid X_1^{\left( 1,2\right) }\rangle $, and particles 3 and 4 are in the
entangled state $\mid X_1^{\left( 3,4\right) }\rangle $. Therefore the total
state is 
\begin{eqnarray}
&\mid &\Psi _{1234}\rangle =\mid X_2^{\left( 1,2\right) }\rangle \mid
X_2^{\left( 3,4\right) }\rangle =\frac 14\left( \mid 1_1\rangle \mid
0_2\rangle +\mid 2_1\rangle \mid 1_2\rangle -\mid 3_1\rangle \mid 2_2\rangle
-\mid 0_1\rangle \mid 3_2\rangle \right)   \nonumber \\
&&\otimes \left( \mid 1_3\rangle \mid 0_4\rangle +\mid 2_3\rangle \mid
1_4\rangle -\mid 3_3\rangle \mid 2_4\rangle -\mid 0_3\rangle \mid 3_4\rangle
\right)   \nonumber \\
&=&\frac 14\left( \mid 1_1\rangle \mid 0_2\rangle \mid 1_3\rangle \mid
0_4\rangle +\mid 1_1\rangle \mid 0_2\rangle \mid 2_3\rangle \mid 1_4\rangle
-\cdots +\mid 0_1\rangle \mid 3_2\rangle \mid 0_3\rangle \mid 3_4\rangle
\right)   \nonumber \\
&=&\frac 18\left\{ 
\begin{array}{c}
\mid 1_1\rangle \left( \mid W_3^{\left( 2,3\right) }\rangle -\mid
X_3^{\left( 2,3\right) }\rangle -\mid Y_3^{\left( 2,3\right) }\rangle +\mid
Z_3^{\left( 2,3\right) }\rangle \right) \mid 0_4\rangle  \\ 
+\mid 1_1\rangle \left( \mid W_2^{\left( 2,3\right) }\rangle -\mid
X_2^{\left( 2,3\right) }\rangle -\mid Y_2^{\left( 2,3\right) }\rangle +\mid
Z_2^{\left( 2,3\right) }\rangle \right) \mid 1_4\rangle  \\ 
-\cdots +\mid 0_1\rangle \left( \mid W_3^{\left( 2,3\right) }\rangle +\mid
X_3^{\left( 2,3\right) }\rangle +\mid Y_3^{\left( 2,3\right) }\rangle +\mid
Z_3^{\left( 2,3\right) }\rangle \right) \mid 3_4\rangle 
\end{array}
\right\} 
\end{eqnarray}
For $\mid 1_1\rangle \mid 0_4\rangle $, $\mid 1_1\rangle \mid 1_4\rangle
,\cdots ,\mid 0_1\rangle \mid 3_4\rangle $ we use Eq.($2$), and we rewrite $%
\mid \Psi _{1234}\rangle $ in $\left( H_2\otimes H_3\right) \otimes \left(
H_1\otimes H_4\right) $, to obtain 
\begin{equation}
\mid \Psi _{1234}\rangle =\frac 14\left\{ 
\begin{array}{c}
\mid W_0^{\left( 2,3\right) }\rangle \mid Z_2^{\left( 1,4\right) }\rangle
-\mid W_1^{\left( 2,3\right) }\rangle \mid X_3^{\left( 1,4\right) }\rangle 
\\ 
-\mid W_2^{\left( 2,,3\right) }\rangle \mid Y_0^{\left( 1,4\right) }\rangle
+\mid W_3^{\left( 2,3\right) }\rangle \mid W_1^{\left( 1,4\right) }\rangle 
\\ 
+\mid X_0^{\left( 2,3\right) }\rangle \mid X_2^{\left( 1,4\right) }\rangle
-\mid X_1^{\left( 2,3\right) }\rangle \mid X_3^{\left( 1,4\right) }\rangle 
\\ 
-\mid X_2^{\left( 2,3\right) }\rangle \mid X_0^{\left( 1,4\right) }\rangle
-\mid X_3^{\left( 2,3\right) }\rangle \mid X_1^{\left( 1,4\right) }\rangle 
\\ 
-\mid Y_0^{\left( 2,3\right) }\rangle \mid W_2^{\left( 1,4\right) }\rangle
+\mid Y_1^{\left( 2,3\right) }\rangle \mid Y_3^{\left( 1,4\right) }\rangle 
\\ 
+\mid Y_2^{\left( 2,3\right) }\rangle \mid Z_0^{\left( 1,4\right) }\rangle
-\mid Y_3^{\left( 2,3\right) }\rangle \mid Z_1^{\left( 1,4\right) }\rangle 
\\ 
-\mid Z_0^{\left( 2,3\right) }\rangle \mid Y_2^{\left( 1,4\right) }\rangle
-\mid Z_1^{\left( 2,3\right) }\rangle \mid Z_3^{\left( 1,4\right) }\rangle 
\\ 
+\mid Z_2^{\left( 2,3\right) }\rangle \mid W_0^{\left( 1,4\right) }\rangle
+\mid Z_3^{\left( 2,3\right) }\rangle \mid Z_1^{\left( 1,4\right) }\rangle 
\end{array}
\right\} 
\end{equation}
This means that when Bob makes a joint measurement of particles 2 and 3, so
that the wave function $\mid \Psi _{1234}\rangle $ collapses to only one of
the above 16 states (say, $\mid W_1^{\left( 2,3\right) }\rangle $ ) with
probability $\frac 1{16}$, then there appears one corresponding entanglement
(say, $\mid X_3^{\left( 1,4\right) }\rangle )$ between particles 1 and 4,
etc. To sum up, the four-level quantum entanglement swapping can be realized
by this way.

As an application of the above schemes, we prove that there may be some
collective translations of multipartite quantum entanglement. Here we need
to use the concepts of the UPB[12-14] and the EEB[17]. Consider an M-partite
quantum system $H=\otimes _{i=1}^MH_i,$each $H_i$\ is $d$-dimensional, the
total dimensionality of $H$ is $N=d^M$. An UPB of $H$ is a product basis $%
S=\left\{ \mid \psi _0\rangle ,\cdots ,\mid \psi _{m-1}\rangle \right\} $,
which spans a subspace $H_S$ in $H$, and the complementary subspace $H{\cal -%
}H_S$ contains no product state. It is known[12,14] that $m\geqslant $ $%
M\left( d-1\right) +1.$ Following [17], if $T=\left\{ \mid \varepsilon
_0\rangle ,\mid \varepsilon _2\rangle ,\cdots ,\mid \varepsilon
_{n-1}\rangle \right\} \left( m+n=d^M\right) $ is a set of entangled
pure-states, and $B=S\cup T$ forms an orthogonal complete basis of $H,$ then
we call $T$ an EEB. Obviously, an arbitrary linear combination of $\mid
\varepsilon _0\rangle ,\mid \varepsilon _2\rangle ,\cdots ,\mid \varepsilon
_{n-1}\rangle $ still is an entangled state in $H$. We call the subspace $%
H_{EES}$ spanned by $T$ the exact entanglement space (EES), in [17] we have
proved the existence of $H_{EES}$. Evidently $H=H_S\oplus H_{EES}$.

Now we take $m=$ $M\left( d-1\right) +1$ and consider the positive integer
solutions $d$ of the following equation 
\begin{equation}
d^M-m=d^M-M\left( d-1\right) -1=4
\end{equation}
There are two solutions, i.e. $M=2,\;d=3$ and $M=3,\;d=2.$ This means that
in a bipartite qutrit system or in a tripartite qubit system, there is a
four dimensional EES (for concrete examples, see [17]). Here for each cases,
we denote the EEB by $\left\{ \mid \varepsilon _0\rangle ,\mid \varepsilon
_1\rangle ,\mid \varepsilon _2\rangle ,\mid \varepsilon _3\rangle \right\} ,$
which spans the entanglement space $H_{EES}.$

Now, we consider a special entanglement problem as follows. The Hilbert
space ${\cal H}{\Bbb =}H_{EES}^{\left( 1\right) }\otimes H_{EES}^{\left(
2\right) }$ has an orthogonal complete basis $\left\{ \mid \varepsilon
_i^{\left( 1\right) }\rangle \mid \varepsilon _j^{\left( 2\right) }\rangle
\right\} \left( i,j=0,1,2,3\right) $. In ${\cal H}$ the general form of a
pure-state $\mid \Phi \rangle $\ is 
\begin{equation}
\mid \Phi \rangle =\sum_{i,j=0}^3f_{ij}\mid \varepsilon _i^{\left( 1\right)
}\rangle \mid \varepsilon _j^{\left( 2\right) }\rangle
\end{equation}
We define a pure-state $\mid \Phi \rangle \in {\cal H}$\ to be `separable in 
${\cal H}$' if and only if it can be decomposed as $\mid \Phi \rangle =\mid
\Phi _1\rangle \mid \Phi _2\rangle ,\;\;\Phi _i\rangle \in H_{EES}^{\left(
i\right) }\left( i=1,2\right) ;$\ Conversely, $\mid \Phi \rangle $\ is
called `entangled in ${\cal H}$'. Notice that since $\mid \varepsilon
_i^{\left( 1\right) }\rangle ,\mid \varepsilon _j^{\left( 2\right) }\rangle $%
\ both are entangled states of $H$\ themselves$,$\ the entanglement in the
above definition, in fact, is a special entanglement, i.e. `entanglement of
entanglement'. Let us make the correspondence 
\begin{equation}
\mid \varepsilon _i^{\left( k\right) }\rangle \rightleftarrows \mid
i_k\rangle \left( i=0,1,2,3;\;k=1,2\right)
\end{equation}
i.e. we regard every basis state $\mid i_k\rangle \left( k=1,2\right) $ an
entangled state in $H_{EES}^{\left( k\right) }\subset H\left( k=1,2\right) ,$
then the entanglement problems of ${\cal H}^{\left( 1,2\right)
}=H_{EES}^{\left( 1\right) }\otimes H_{EES}^{\left( 2\right) }$ are just a
ququadrit entanglement problems, and we can use the basis as $\left\{ \mid
W_i^{\left( 1,2\right) }\rangle ,\mid X_i^{\left( 1,2\right) }\rangle ,\mid
Y_i^{\left( 1,2\right) }\rangle ,\mid Z_i^{\left( 1,2\right) }\rangle
\right\} \left( i=0,1,2,3\right) $, etc..

Now for a tripartite qubit system, suppose that Alice holds three particles $%
1,2,3$ which are in an unknown state $\mid \phi ^{\left( 123\right) }\rangle
=\alpha \mid \varepsilon _0^{\left( 123\right) }\rangle +\beta \mid
\varepsilon _1^{\left( 123\right) }\rangle +\gamma \mid \varepsilon
_2^{\left( 123\right) }\rangle +\delta \mid \varepsilon _3^{\left(
123\right) }\rangle ,$ Bob holds six particles $4,5,\cdots ,9$ which are in
the entangled state $\mid X_1^{\left( 456,789\right) }\rangle =\frac
12\left( 
\begin{array}{c}
\mid \varepsilon _1^{\left( 456\right) }\rangle \mid \varepsilon _0^{\left(
789\right) }\rangle +\mid \varepsilon _2^{\left( 456\right) }\rangle \mid
\varepsilon _1^{\left( 789\right) }\rangle \\ 
-\mid \varepsilon _3^{\left( 456\right) }\rangle \mid \varepsilon _2^{\left(
789\right) }\rangle -\mid \varepsilon _0^{\left( 456\right) }\rangle \mid
\varepsilon _3^{\left( 789\right) }\rangle
\end{array}
\right) ,$ then, as above, we can complete the quantum teleportation of $%
\mid \phi ^{\left( 123\right) }\rangle $ from Alice to remote Clara.

Similarly, suppose that there are twelve spin-$\frac 12$ particles $%
1,2,\cdots 6,1^{\prime },2^{\prime },\cdots 6^{\prime },$ Alice holds
particles $1,2,3$, Bob holds six particles $4,5,6,1^{\prime },2^{\prime
},3^{\prime }$, Clara holds particles $4^{\prime },5^{\prime },6^{\prime },$
particles $1,2,3$ and $4,5,6$ are in the entangled state $\mid X_1^{\left(
123,456\right) }\rangle $, and particles $1^{\prime },2^{\prime },3^{\prime
} $ and $4^{\prime },5^{\prime },6^{\prime }$ are in the entangled state $%
\mid X_1^{\left( 1^{\prime }2^{\prime }3^{\prime },4^{\prime }5^{\prime
}6^{\prime }\right) }\rangle $. Therefore the total state is $\mid \Psi
_{1234}\rangle =\mid X_2^{\left( 123,456\right) }\rangle \mid X_2^{\left(
1^{\prime }2^{\prime }3^{\prime },4^{\prime },5^{\prime },6^{\prime }\right)
}\rangle ,$ and according to the above steps of swapping, it can be swapped
into the entanglement between particle groups $(1,2,3)$ and ($4^{\prime
},5^{\prime },6^{\prime }$), and between groups $(4,5,6)$ and ($1^{\prime
},2^{\prime },3^{\prime }$), etc.. The case of a bipartite qutrit systems is
similar.

Obviously, the above quantum teleportation and swapping, in fact, are some
collective translations of quantum entanglement. Especially, the foundations
of this method are the four-level quantum teleportation, swapping, and EEB.

At last, we briefly mention the problem of generalization. We find that if
the above scheme is directly used to the three-level systems, then some
information will be lost. Next, the schemes also can be extended to the
N-level (N$\geqslant 5)$\ systems, however the results are quite complex.

{\bf Conclusion: }In the ququadrit systems there are special bases. Using
these bases, we can simply realize the four-level quantum teleportation and
swapping. As an application of the above schemes, it is proved that in the $%
2\times 2\times 2$ or $3\times 3$ systems there may be some collective
translations of multipartite quantum entanglement.


\begin{thebibliography}{99}
\bibitem{}  C. H. Bennett, G. Brassard, C. Cr\'{e}peau, R. Jozsa, A. Peres,
and W. K. Wootters, Phys. Rev. Lett., {\bf 70}(1993)1895.

\bibitem{}  M. A. Nielsen and I. L. Chuang, {\it Quantum Computation and
Quantum Information.} Cambridge University Press (2000).

\bibitem{}  M. Zukowski, A. Zeilinger, M. A. Horne, and A. Ekert, Phys. Rev.
Lett., {\bf 71}(1993)4287.

\bibitem{}  F. Verstraete and H. Verschelde, Phys. Rev. Lett., {\bf 90}%
(2003)097901.

\bibitem{}  J. D. Zhou, G. Hou, and Y. D. Zhang, Phys. Rev. A, {\bf 64}%
(2001)012301

\bibitem{}  W. Son, J. H. Lee, M. S. Kim, and Y. J. Park, Phys. Rev. A, {\bf %
64}(2001)064304.

\bibitem{}  J. H. Lee, H. Min, and S. D. Oh, Phys. Rev. A, {\bf 66}%
(2002)052318.

\bibitem{}  A. Grudka, Acta Phys. Slov., {\bf 54}(2004)91.

\bibitem{}  A. Grudka and R. W. Chhajlany, Acta Phys. Pol., {\bf A 104}%
(2003)409.

\bibitem{}  G. Rigolin, Phys. Rev. A, {\bf 71}(2005)032303.

\bibitem{}  X. H. Ge and Y. G. Shen, Phys. Lett. B{\bf 606} (2005)184.

\bibitem{}  C. H. Bennett , D. P. DiVincenzo , T. Mor, P. W. Shor, J. A.
Smolin , and B. M. Terhal, Phys. Rev. Lett., {\bf 82}(1999)5385.

\bibitem{}  B. M. Terhal, Lin. Alg. Appl., {\bf 323}(2000)61.

\bibitem{}  D. P. DiVincenzo , T. Mor, P. W. Shor, J. A. Smolin and B. M.
Terhal, Comm. Math. Phys., {\bf 238}(2003)379.

\bibitem{}  S. Chaturvedi, Phys. Rev. A, 65(2002)042322.

\bibitem{}  A. O. Pittenger, Linear Algebr. Appl., {\bf 359}(2003)235.

\bibitem{}  Z. Z. Zhong, Phys. Rev. A, {\bf 70}(2004)044302.
\end{thebibliography}
\end{document}